\documentclass[aps,twocolumn,amsfonts,amssymb,superscriptaddress]{revtex4-1}
\usepackage{graphicx}
\usepackage{epstopdf} 
\usepackage{dcolumn}
\usepackage{bm}
\usepackage{bbm,bbold}
\usepackage{color}
\usepackage{xcolor, soul}
\sethlcolor{green}
\usepackage{amsfonts}
\usepackage{amsmath}
\usepackage{mdframed}
\usepackage[normalem]{ulem}
\usepackage{mathrsfs}   
\usepackage[none]{hyphenat}
\usepackage{subfigure}
\usepackage{float}
\usepackage[colorlinks=true,citecolor=blue]{hyperref}
\hypersetup{colorlinks=true,citecolor=blue,linkcolor=blue,urlcolor=blue}
\DeclareRobustCommand{\rchi}{{\mathpalette\irchi\relax}}
\newcommand{\irchi}[2]{\raisebox{\depth}{$#1\chi$}} 

\begin{document}
	\title{Second Harmonic Helicity and Faraday Rotation in Gated Single-Layer 1T$'$-WTe$_2$}
	\author{Pankaj Bhalla}
	\affiliation{Nordita, KTH Royal Institute of Technology and Stockholm University,
     Hannes Alfvéns väg 12, 114 21 Stockholm, Sweden}
    \author{Habib Rostami}
	\affiliation{Nordita, KTH Royal Institute of Technology and Stockholm University, 
	 Hannes Alfvéns väg 12, 114 21 Stockholm, Sweden}

\date{\today}

\begin{abstract}
Single-layer of 1T$'$ phase of WTe$_2$ provides a rich platform for exotic physical properties such as nonlinear Hall effect and high-temperature quantum spin hall transport. Utilizing a continuum model and diagrammatic method, we calculate the second harmonic conductivity of monolayer 1T$'$-WTe$_2$ modulated by an external vertical electric field and electron doping. We obtain a finite helicity and Faraday rotation for the second harmonic signal in response to linearly polarized incident light in the presence of time-reversal symmetry. The second harmonic signal's helicity is highly controllable by altering the bias potential and serves as an optical indicator of the nonlinear Hall current. Our study motivates future experimental investigation for the helicity spectroscopy of two dimensional materials. 
\end{abstract}

\maketitle

\section{Introduction}
Two dimensional (2D) materials such as transition metal dichalcogenides (TMDs), with chemical formula MX$_2$ where M stands for the transition metal atom (W, Mo) and X is the chalcogen atom (Te, Se or S), appear in different crystalline structures such as hexagonal 2H, tetragonal 1T, and distorted 1T$'$ and 1T$_d$ structures \cite{joy_JACS1999, eda_ACSN2012, qian_Sc2014}. Noncentrosymmetric TMDs exhibit unique physical properties such as circular dichroism, piezoelectricity, nonlinear Hall effect and second harmonic generation due to their distinct phases \cite{mak_NN2012, duerloo_JPCL2012, shirodkar_PRL2014, jiang_PRL2015, bruyer_PRB2016, morozovska_PRB2020, kawaguchi_NC2021, khan_AFM2021, Shao_PNAS2021}.
The distorted 2D WTe$_2$ is attracting a surge of interest due to diverse ground state phases such as quantum spin Hall, superconductivity, polar metal and ferroelectricity \cite{tang_NP2017, garcia_PRL2020, fei_NP2018, huang_ACSN2018, sharma_SA2019}. 

The 1T phase of WTe$_2$ has a rhombohedral (ABC) stacking with one tungsten layer sandwiched between two tellurium layers and has $D_{3d}$ point group \cite{yang_NP2017}. However, the free-standing 1T system is unstable and it undergoes a spontaneous lattice distortion to form a period-doubling distorted structure known as 1T$'$ phase having point group $C_{2h}$ \cite{Li_JPCM2020}. Note that, the metal atoms in 1T' phase rearrange in the zigzag chain. The 1T$'$ phase possesses mirror plane symmetry $\mathcal{M}_x$ perpendicular to $x$-direction and two-fold rotational symmetry $C_{2x}$, hence remains $\mathcal{I} = \mathcal{M}_xC_{2x}$ inversion symmetric \cite{jia_PRB2017, fei_NP2017, tang_NP2017}.
Moreover, the monolayer 1T$_d$ structure, with point group $C_{1s}$ \cite{xu_NP2018}, breaks the two-fold rotational symmetry due to further distortion than the 1T$'$ phase and preserves the mirror symmetry $\mathcal{M}_x$, hence it breaks the inversion symmetry \cite{xu_NP2018}. As a result of this inversion symmetry breaking, a weak intrinsic in-plane displacement field emerges in 1T$_d$ structure and the dispersion is spin-polarized.

\begin{figure}[t]
    \centering
    \includegraphics[width=8.5cm]{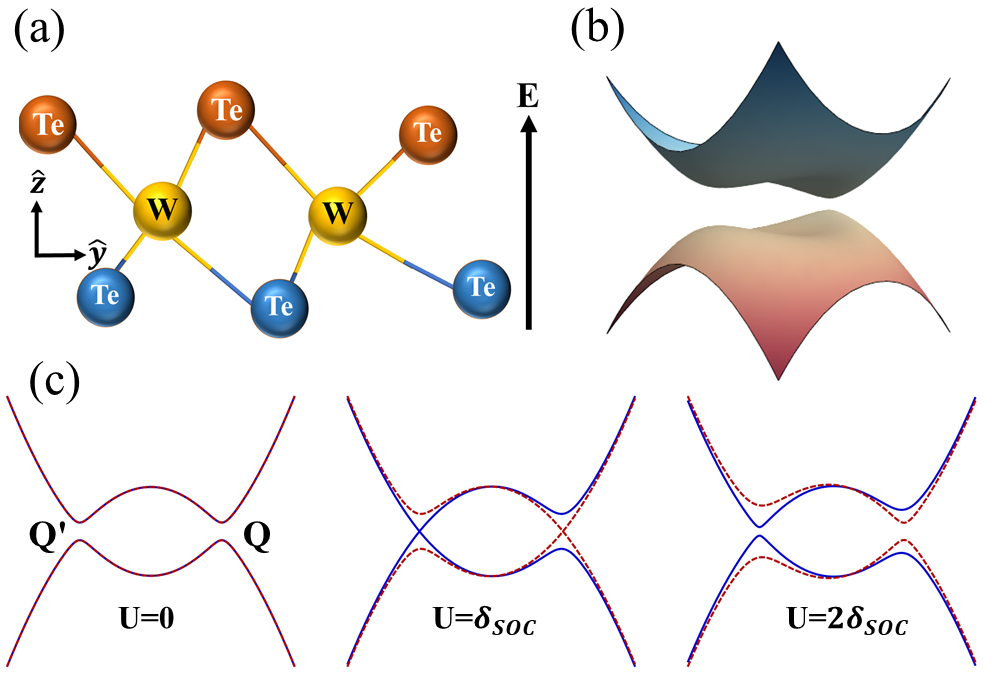}
    \caption{(a): Lattice distorted 1T$'$ structure of WTe$_2$ in the presence of an external vertical electric field ${\bm E}$ where top and bottom atomic layers are shown in different colours, (b): Energy dispersion of single-layer 1T$'$-WTe$_2$ in the momentum space with the finite bias potential $U$, (c): Dispersion along $k_x$ direction at different bias potential $U=0$, $U=\delta_{\rm SOC}$, and $U=1.5\delta_{\rm SOC}$ having blue solid curve for the spin index $s = +1$ and red dashed for $s =-1$. Here $Q' = -Q$, and $\delta_{\rm SOC}= v_x Q$, where $v_x$ is the component of the velocity along $x$-direction.}
    \label{fig:Fig1}
\end{figure}

By applying an external vertical electric field in a dual-gated system of distorted (1T$'$ and 1T$_d$) WTe$_2$, one can efficiently create a controllable displacement field \cite{xu_NP2018} similar to the case of a biased bilayer graphene \cite{min_PRB2007, oostinga_NM2008, zhang_N2009, tay_PRL2010} -- see Fig.~\ref{fig:Fig1}a. Upon the usage of a strong external electric field, the field-induced bandgap dominates the intrinsic one in the 1T$_d$ phase.
In the absence of the spin-orbit interaction and the vertical field, the low energy dispersion of the WTe$_2$ reveals two topologically protected tilted Dirac cones at the neighbourhood of the $\Gamma$ point \cite{lukas_PRX2016, shi_PRB2019, lau_PRM2019}.
The spin-orbit induced band-inversion drives the system into a quantum spin Hall phase in a reasonably high temperature 100K \cite{duerloo_NC2014, choe_PRB2016, fei_NP2017, ugenda_NC2018}. 
The schematic view of the 1T$'$-WTe$_2$ dispersion is shown in Fig.~\ref{fig:Fig1}. The energy bands are degenerate at zero bias potential with a band gap due to the spin-orbit interaction. However, the band degeneracy breaks down by turning on the bias potential $U$. Interestingly, on approaching $U$ to the strength of the spin-orbit coupling ($\delta_{{\rm SOC}}$) the spin-polarized band gap vanishes around one Dirac point. With more distortion ($U > \delta_{{\rm SOC}}$), the bands remain non-degenerate and the gap again opens up as shown in Fig.~\ref{fig:Fig1}c.

Strong and diverse forms of nonlinear response in 2D materials are drawing attention for applications in all-optical modulators \cite{wang_NN2012, mueller_NPJ2D2018,soavi_nn_2018,Wang_AdM_2021,klimmer_NP2021}. Especially, a strong nonlinear Hall response has been measured in single and bilayer of WTe$_2$ that can be described in terms of the Fermi surface average of the Berry curvature derivative, the so called Berry curvature dipole \cite{sodemann_PRL2015, zhang_PRB2018, you_PRB2018, quereda_NC2018, facio_PRL2018,fleisher_NP2014, fernando_NC2017,ma_NP2017, vander_PRB2017, sabbaghi_PRB2018, zhang_PRB2018, habib_PRB2018, juan_PRR2020, matsyshyn_PRL2019, li_NM2019, shi_SA2019, bhalla_PRL2020, bhalla_PRB2021}. The second harmonic generation (SHG) is utilized to measure the internal lattice distortion (strain) in 2D crystals \cite{mennel_NC2018}. The dynamical form of the nonlinear Hall effect manifests in the second harmonic transverse current in response to the linearly polarized light in noncentrosymmetric 2D materials such as biased single and few layer WTe$_2$ \cite{moore_prl2010, hosur_prb2011, ma_Nat2019, du_PRL2018, kang_NM2019, bhalla_arxiv2021}. Interestingly, theoretical studies provide the direct correspondence of high harmonic helicity (circular dichroism) and topological nature of the electronic band structure \cite{zhang_PRA2016, zhang_PRB2019}. For instance, it has been shown that the nonlinear helicity spectroscopy can sharply distinguish between trivial and topological phases of the Haldane model in the hexagonal 2D materials \cite{chacon_PRB2020, silva_NP2019}. 

Despite extensive studies of all-optical amplitude modulation, the phase and polarization modulation of the second harmonic signal are not systematically studied in TMD materials. Faraday rotation and dichroism effects can be employed to explore the polarization of second harmonic radiation. The Faraday effect is the rotation of the polarization plane of a linearly polarized light upon its propagation through a medium \cite{faraday_PTRSL1846, stern_PR1964, connell_PRB1982, boyd_book, budker_RMP2002, bermen_AJP2010, bloembergen_book} which decomposes the linear polarized light into left and right-handed circular components. In isotropic materials, the Faraday rotation emerges only after breaking the time-reversal symmetry. In such cases, the left and right-handed components of the polarized light experience different refraction index and hence propagate with different phase velocities that lead to the rotation of a polarization plane (Faraday rotation) as well as finite ellipticity (helicity). The nonlinear hall effect does not require time-reversal symmetry to be broken and therefore a finite helicity and Faraday rotation are expected for the SHG signal. To the best of our knowledge, the SHG signal's polarization in single-layer 1T$'$-WTe$_2$ has not been explored microscopically.  

This study aims to fill this gap by developing microscopic continuum analysis of the polarization and amplitude modulations in single-layer 1T$'$-WTe$_2$. We study the second harmonic response for the time-reversal symmetric and distorted 1T$'$-WTe$_2$ in the presence of a vertical electric field. We first propose the theory for the nonlinear Faraday rotation angle and the second harmonic helicity in terms of the second-order optical conductivity.
We explore the effects of bias potential induced by the vertical  electric field on the second harmonic susceptibility, Faraday rotation angle, and second harmonic helicity. We find that significantly large helicity can be achieved near the interband transitions. A linear light with frequency $\omega$ can be effectively converted to a circularly polarized light with frequency $2\omega$. The process is controllable by the interplay of the Fermi energy and the bias potential. 

\section{Theoretical Method}
We consider a low energy model in the 2D momentum space for the distorted 1T$'$ phase of WTe$_2$ which is described as \cite{xu_NP2018}
\begin{equation} \label{eqn:Ham}
    \mathcal{\hat H} = A k^2  {\hat I} + (\delta + Bk^2) \hat \sigma_z + \hbar v_y k_y \hat \sigma_y 
    + (U + v_x s k_x) \hat \sigma_x,
\end{equation}
where $s=\pm1$ stands for the spin index. The magnetization axis is assumed along $\hat{y}$-direction and the spin-mixing term $ U'\hat\sigma_y \hat s_x$ is neglected since $U\gg U'$ \cite{xu_NP2018}. 
The Pauli matrix $\hat \sigma_{x,y,z}$ is represented in the orbital basis $\{\psi_{+},\psi_{-}\}=$ $\{$p-orbital of Te, d-orbital of W$\}$ which are separated by the fundamental gap $2\delta$ at $\Gamma$ point and $\hat{I}$ refers to $2\times 2$ identity matrix. 
Here, $U$ represents the coupling between the out-of-plane electric field and the orbitals, the wave vector ${\bm k} = (k_x,k_y)$ having $k = |{\bm k}|$. The ${\bm k}\cdot {\bm p}$ parameters $v_x$, $v_y$, $A$ and $B$ are obtained after fitting to realistic low energy dispersion of 1T$'$-WTe$_2$ \cite{papaj_PRL2019, xu_NP2018}. 
In our analysis, we focus on the impact of bias potential on response.  
Corresponding to the Hamiltonian Eq.~\eqref{eqn:Ham}, the energy dispersion is given by  
\begin{align}
\varepsilon_{\bm {k},s}^{\lambda} =A k^2 +\lambda\sqrt{(\delta + Bk^2)^2 + \hbar^2 v_y^2k_y^2 + (U + s v_x k_x)^2},
\end{align}
and the eigen vectors are $|u_{\bm {k},s}^\lambda \rangle = 1/\sqrt{2}[\sqrt{1\pm b_{k,s}}, \sqrt{1 \mp b_{k,s}}]$, where $b_{k,s} = (\delta + B k^2)/E_{\bm {k},s}$ having $E_{{\bm k},s} = [(\delta + B k^2)^2 + \hbar^2 v_y^2k_y^2 + (U + s v_x k_x)^2]^{1/2}$, with $\lambda = + (-)$ for the conduction (valence) band. In addition to the time-reversal symmetry, this model also possesses the mirror symmetry along $\hat{x}$-direction which ensures the dispersion $\varepsilon(k_x,k_y) = \varepsilon(-k_x,k_y)$. 
In the absence of bias potential $U=0$ and for $\delta/B<0$, the system has two gapless nodes (valleys) at $\tau {\bm Q} =(\tau \sqrt{|\delta/B|},0)$ with $\tau=\pm$ as the valley index. The schematic picture of the dispersion is shown in Fig.~\ref{fig:Fig1}. The energy gap at valley point $\tau {\bf Q}$ is given by $\Delta_{\pm} = U \pm v_x Q$ for each spin-valley index $\tau s=\pm$. Similar to the Kane-Mele model in graphene with spin-orbit coupling \cite{KM_2005}, the 1T$'$-WTe$_2$ is in the topological (i.e. quantum spin Hall) phase when $|v_x Q|>|U|$ and is trivial for $|v_x Q|<|U|$. The finite bias potential $U$ breaks the inversion symmetry of the system and it opens the gap around two points ($Q$ points) in the Brillouin zone as depicted in Fig.~\ref{fig:Fig1}. However, the time-reversal symmetry of the system remains preserved. Further, this externally tunable bias potential is an artifact of the in-plane electrical polarization induced by an external out-of-plane electric field due to the intrinsic lattice distortion as displayed in Fig.~\ref{fig:Fig1}(a) \cite{xu_NP2018}.

\subsection{Nonlinear response of gated 1T$'$-WTe$_2$}
\begin{figure}[t]
    \centering
    \includegraphics[width=8cm]{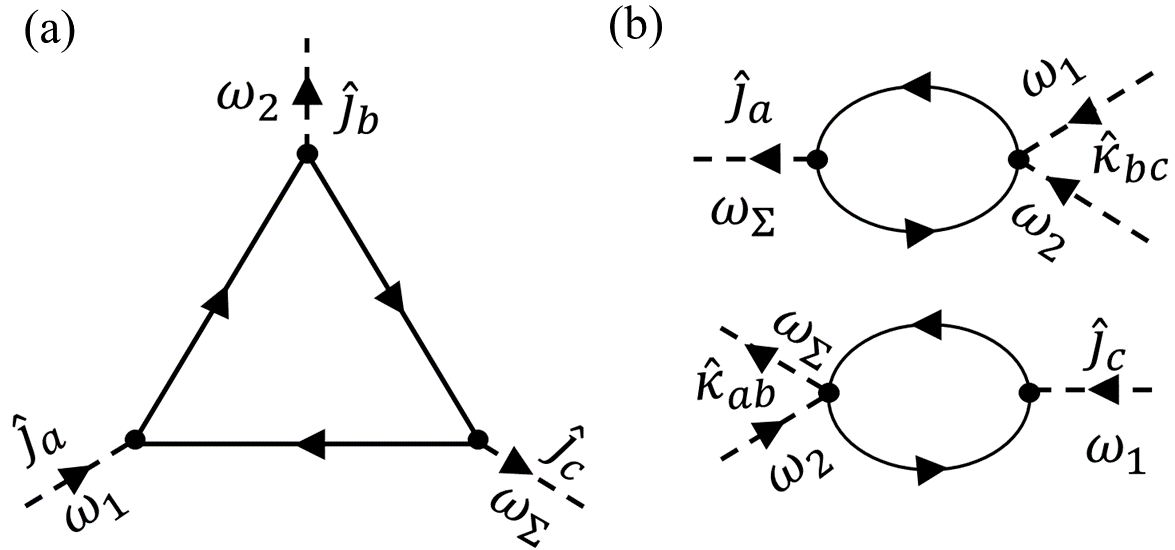}
    \caption{Feynmann diagrams for the (a) paramagnetic and (b) diamagnetic contributions to the second-order response. Here solid lines indicate the electron propagators, and dotted lines refer to external photons. The symbols $\omega_i$ represents the incoming and outgoing frequencies, $\omega_\Sigma = \omega_2+\omega_2$, $\hat{j}_i$ and $\kappa_i$ denote one and two-photon current vertices respectively.}
    \label{fig:Fig2}
\end{figure}

Light-matter interaction is modeled via minimal coupling $\hbar {\bm k}\to \hbar{\bm k} + e {\bm A}(t)$, where ${\bm A(t)}$ is an external vector potential. Due to the nonlinear dispersion of the Hamiltonian, both one and two-photon couplings are present \cite{rostami_AP_2021}. 
We begin with the phenomenological relation of the second-order current in the frequency domain in response to an external vector potential $\bm{A}(t)$, with the time dependent and spatial homogeneous electric field $\bm{E}(t) = -\partial_t \bm{A}(t)$ \cite{boyd_book}
\begin{align} \label{eqn:current}
    J_a^{(2)}(\omega_\Sigma) &= \sum_{\omega_1,\omega_2} \sum_{bc} \rchi_{abc}^{(2)}(\omega_\Sigma; \omega_1,\omega_2)A_b(\omega_1)A_c(\omega_2)
    \nonumber\\&\times \delta(\omega_\Sigma - \omega_1 -\omega_2).
\end{align}
Here, we define $\bm{A}(\omega) = -i\bm{E}(\omega)/\omega$. For the sake of convenience, we write the tensor quantity as $\rchi_{abc}^{(2)}(\omega_1,\omega_2)$ in further calculations. The latter second-order response tensor includes two terms \cite{Antti_nc_2017,rostami_AP_2021}:
\begin{align} \label{eqn:chi}
    \rchi_{abc}^{(2)}(\omega_1,\omega_2) = \rchi_{abc}^{(2),D}(\omega_1,\omega_2)+\rchi_{abc}^{(2),P}(\omega_1,\omega_2),
\end{align}
where the terms $\rchi_{abc}^{(2),D}$ and $\rchi_{abc}^{(2),P}$ refer to the diamagnetic and paramagnetic contributions respectively. 
The paramagnetic susceptibility, a three-point retarded correlation function of current operator components is defined diagrammatically in Fig. \ref{fig:Fig2}a.  

Using the many-body diagrammatic perturbation theory, the second-order paramagnetic susceptibility after performing the Matsubara frequency summation and then considering the analytic continuation $i\omega_n \to \omega + i\eta$ with $\eta\to 0^+$ can be expressed as \cite{mahan_book, habib_PRR2020} 
\begin{align} \label{eqn:chi_para} \nonumber
    \rchi_{abc}^{(2),P}(\omega_1,\omega_2) = \sum_{\mathcal{P}} \sum_{\{\lambda_i \}} \sum_{\bm{k},s} \frac{j_a^{\lambda_1\lambda_2}j_b^{\lambda_2\lambda_3}j_c^{\lambda_3\lambda_1}}{\hbar\omega_\Sigma + \varepsilon_{\bm{k},s}^{\lambda_2} - \varepsilon_{\bm{k},s}^{\lambda_1} + i\eta}\\
    \times
    \bigg\{ \frac{f(\varepsilon_{\bm{k},s}^{\lambda_2}) -f(\varepsilon_{\bm{k},s}^{\lambda_3}) }{\hbar\omega_1 + \varepsilon_{\bm{k},s}^{\lambda_2} - \varepsilon_{\bm{k},s}^{\lambda_3} + i\eta} - \frac{f(\varepsilon_{\bm{k},s}^{\lambda_3}) -f(\varepsilon_{\bm{k},s}^{\lambda_1})}{\hbar\omega_2 + \varepsilon_{\bm{k},s}^{\lambda_3} - \varepsilon_{\bm{k},s}^{\lambda_1} + i\eta} \bigg\}.
\end{align}
The one-photon coupling is the standard current vertex $\hat j_a =- (e/\hbar) \partial_{k_a} \hat {\cal H}$ which is also known as the paramagnetic current operator.
Note that $ \sum_{\mathcal{P}}$ stands for the intrinsic permutation symmetry $(b,\omega_1) \Longleftrightarrow (c,\omega_2)$. 
Here $f(\varepsilon_{\bm{k},s}^{\lambda}) = [1+e^{\beta(\varepsilon_{\bm{k},s}^{\lambda}-\mu)}]^{-1}$ is the Fermi-Dirac distribution function, $\mu$ is the chemical potential, and $\beta = 1/k_B T$ having $k_B$ as the Boltzmann constant, $T$ as electron temperature. Similarly, the diamagnetic contribution to the second-order response according to the Fig.~\ref{fig:Fig2}b is given by
\begin{align} \label{eqn:chi_dia} \nonumber
    & \rchi_{abc}^{(2),D}(\omega_1,\omega_2) = \sum_{\mathcal{P}} \sum_{\{\lambda_i \}} \sum_{\bm{k},s} \bigg\{ \frac{j_a^{\lambda_1\lambda_2}\kappa_{bc}^{\lambda_2\lambda_1}(f(\varepsilon_{\bm{k},s}^{\lambda_2}) -f(\varepsilon_{\bm{k},s}^{\lambda_1})) }{\hbar\omega_1 + \varepsilon_{\bm{k},s}^{\lambda_1} - \varepsilon_{\bm{k},s}^{\lambda_2} + i\eta}
     \nonumber\\ &
     +
     \frac{1}{2} \frac{(j_b^{\lambda_2\lambda_1}\kappa_{ac}^{\lambda_1\lambda_2}+j_c^{\lambda_2\lambda_1}\kappa_{ba}^{\lambda_1\lambda_2}) (f(\varepsilon_{\bm{k},s}^{\lambda_2}) -f(\varepsilon_{\bm{k},s}^{\lambda_1})) }{\hbar\omega_\Sigma + \varepsilon_{\bm{k},s}^{\lambda_1} - \varepsilon_{\bm{k},s}^{\lambda_2} + i\eta}
     \bigg\}.
\end{align}
 The two-photon coupling is given by $\hat \kappa_{ab} = -(e/\hbar)^2 \partial_{k_a} \partial_{k_b} \hat {\cal H} $, also termed as the diamagnetic current operator in the context of superconductors. Considering the explicit form of the ${\bm k} \cdot {\bm p}$ Hamiltonian Eq.~\eqref{eqn:Ham}, the only non-vanishing components of the two-photon vertex coupling are $\hat \kappa_{xx}$ and $\hat \kappa_{yy}$. 

Due to the mirror symmetry $x\to-x$, the elements of second-order susceptibility $\rchi_{abc}^{(2)}$ with an odd number of $x$ indices will vanish. The remaining non-vanishing components of the second harmonic susceptibility with even number of $x$ indices are $\rchi_{yxx}^{(2)}$, $\rchi_{xyx}^{(2)}$, $\rchi_{xxy}^{(2)}$, and $\rchi_{yyy}^{(2)}$ which we have computed in the next subsection. Additionally, the two components $\rchi_{xxy}^{(2)} = \rchi_{xyx}^{(2)}$ on interchanging the last two spatial indices by symmetry. Hence, one is left with three independent third rank tensor components. 

Using the total second-order susceptibility, one can obtain the second-order conductivity as follows 
\begin{align}
    \sigma_{abc}^{(2)}(\omega_1,\omega_2) = - \frac{\rchi_{abc}^{(2)}(\omega_1,\omega_2)}{\omega_1\omega_2}. 
\end{align}
Here, we consider $\omega_1 = \omega_2 = \omega$ to compute second harmonic conductivities which will be further used to obtain the polarization quantities such as the Faraday rotation and the helicity below.

\begin{figure*}
    \centering
    \includegraphics[width=19cm]{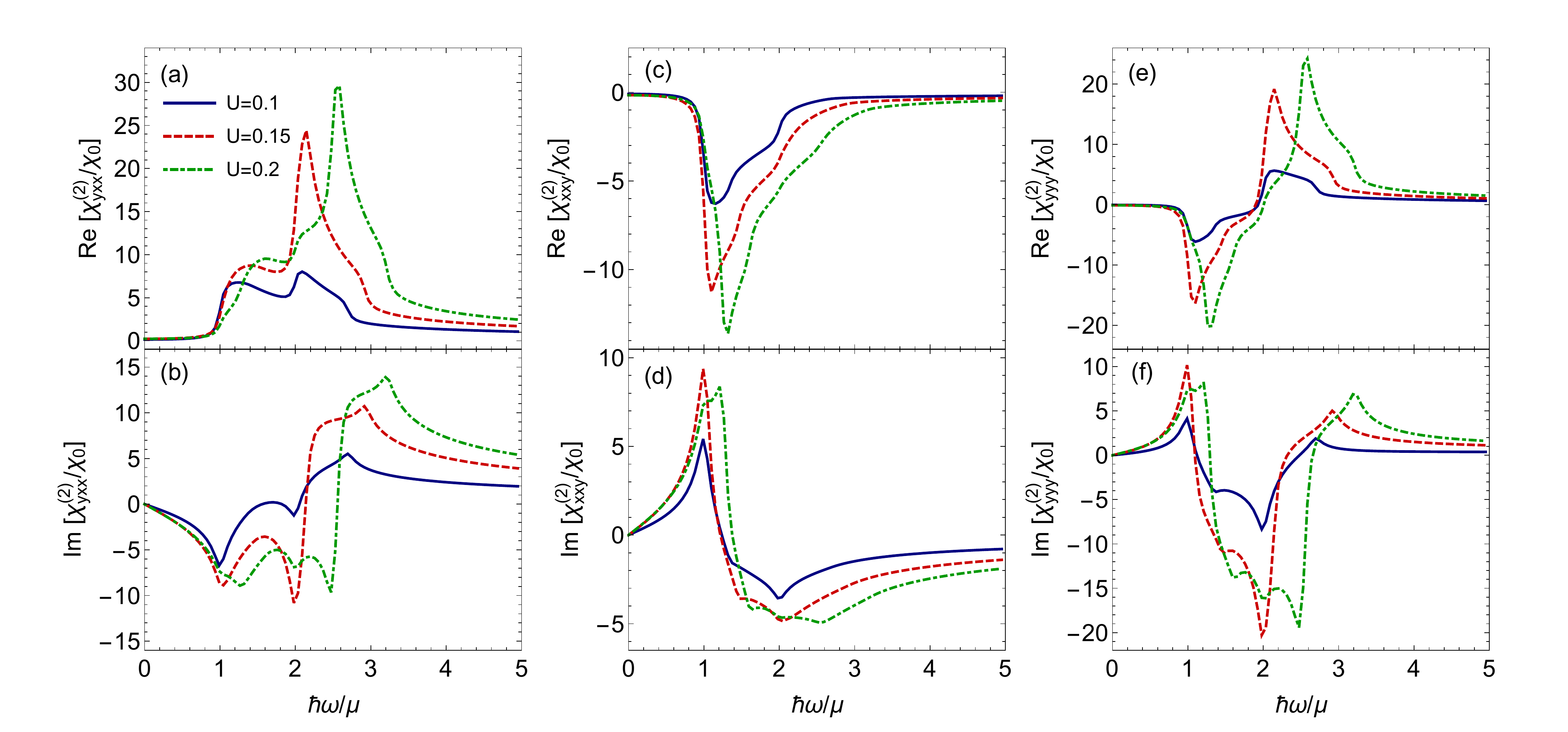}
    \caption{The real and imaginary parts of three distinct components of the second-order susceptibility in response to the linearly polarized beam at three different bias potential values as a function of an incident energy scaled by the chemical potential. Here we set the parameters $A=0$, $\delta = -0.25$ eV, $B = 1$ eV\AA$^{2}$, $v_x = 0.1$ eV\AA, $v_y = 0.3$ eV\AA, and $\mu = 0.2$ eV and $\rchi_0 = -e^3 v_y/(4\pi^2 \hbar^2)$.}
    \label{fig:Fig3}
\end{figure*}

\subsection{Second harmonic signal polarization}
For the analytical derivation in the present study, we assume the linearly polarized light with normal incidence and a generic polarization direction $\bm{E}(t) = \hat{\bm \epsilon}(\phi) E_0 e^{i({\bf q}\cdot r - \omega t)} + c.c.$. Here $E_0$ is the amplitude of the incident beam, $\hat{\bm \epsilon}(\phi) = \hat{x}\cos\phi  + \hat{y}\sin\phi $ is the polarization direction having $\phi$ the polarization angle with the $x$-axis, and the wave vector ${\bf q} = - \hat{\bf z}\omega/c$. At the normal incidence, the second harmonic current (taking $\omega_1 = \omega_2 = \omega$) follows  ${\bf J}^{(2)}   =  E_0^2  (a_{x} \hat{x} + a_{y} \hat {y})$ where the component of the response can be expressed like
\begin{align} \label{eqn:aj}
a_j=\sigma^{(2)}_{jxx} (\cos\phi)^2 + \sigma^{(2)}_{jyy}(\sin\phi)^2 +\sigma^{(2)}_{jxy}  \sin(2\phi).
\end{align}
Here, we have used the symmetry relation $\sigma_{jxy}^{(2)} = \sigma_{jyx}^{(2)}$ in the last term. More specifically, the second harmonic current can be written in terms of the longitudinal (along the polarization direction) and transverse (normal to the polarization direction) components as ${\bf J}^{(2)} =  J^{(2)}_{\rm L} \hat{\bm \epsilon}(\phi) + J^{(2)}_{\rm T} \hat {\bf t}(\phi)$,
where the transverse direction $\hat {\bf t}(\phi) = \hat {\bf z}\times \hat{\bm \epsilon}(\phi)$.
We define the longitudinal and transverse components of the nonlinear conductivity in the form $\sigma^{(2)}_{\rm L} = {J^{(2)}_{\rm L}}/{ E^2_0} = a_x \cos\phi + a_y \sin\phi$ and $\sigma^{(2)}_{\rm T}= {J^{(2)}_{\rm T}}/{E^2_0} = -a_x\sin\phi + a_y\cos\phi$.
Here we consider $a_x = |a_x| e^{i\delta_x}$ and $a_y = |a_y| e^{i\delta_y}$ where $\delta_x$ ($\delta_y$) refers to the phase associated with the $\hat{x}$ ($\hat{y}$) component of the response. 
In the circular basis representation ($\hat{l}$,$\hat{r}$), we can write $a_{r/l} \equiv a_x \pm i a_y$ where $+/-$ sign stands for the right/left-handed circular counterpart. Using the circular decomposition along with the relation between Stokes and polarization parameters \cite{mcmaster_AJP1954, jackson_book, artal_book}, the Faraday rotation angle $\psi$ is defined as
\begin{align}
    \tan (2\psi) = \frac{S_2}{S_1} = - \frac{\text{Im}[a_l^* a_r]}{\text{Re}[a_l^* a_r]}.
\end{align}
Here $S_i$ denotes the Stokes parameters of the monochromatic light \cite{mcmaster_AJP1954} such as $S_0 = |a_l|^2 + |a_r|^2$, $S_1 = 2\text{Re}[a_l^* a_r]$, $S_2 = -2\text{Im}[a_l^* a_r]$, and $S_3 = |a_l|^2 - |a_r|^2$. Similarly, the ellipticity $\rchi$, an angle that defines the amount of elliptic nature from the circular shape is 
\begin{align} \label{eqn:Ellip}
    \tan (2\rchi) = \frac{S_3}{\sqrt{S_1^2 + S_2^2}} = \frac{|a_l|^2 - |a_r|^2}{|a_l^*a_r|}.
\end{align}
\begin{figure*}[t]
    \centering
    \includegraphics[width=17cm]{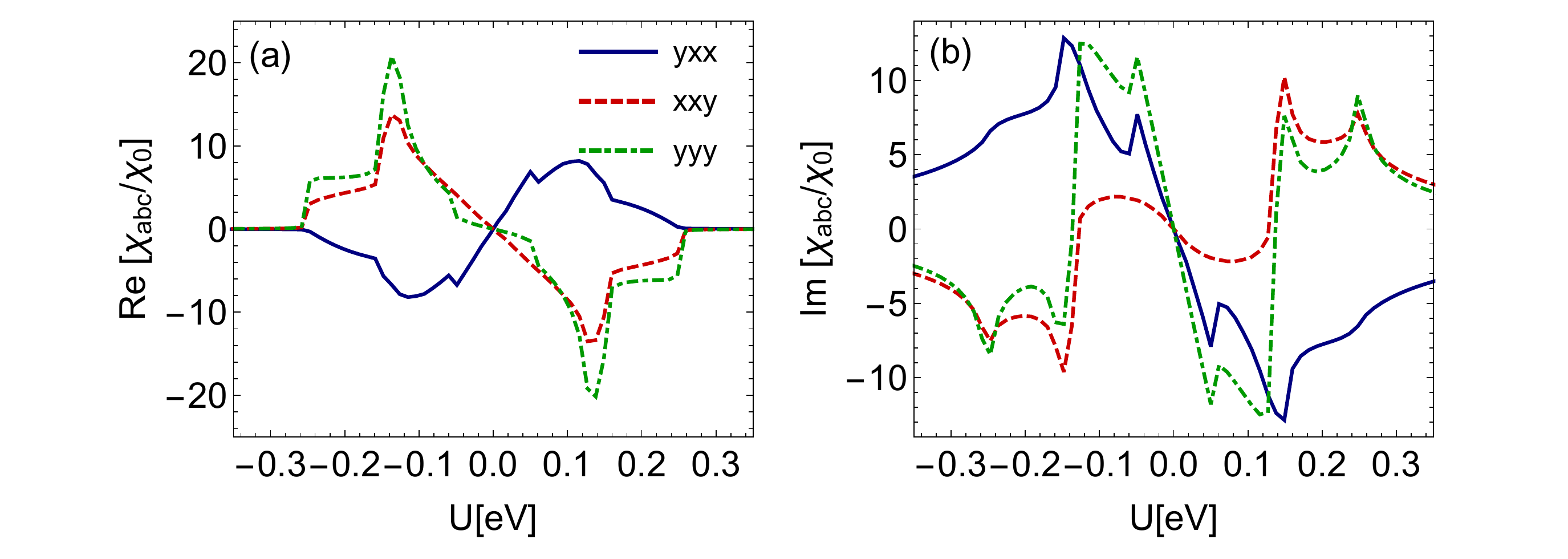}
    \caption{Plot for the different components of the (a) real and (b) imaginary parts of the nonlinear susceptibility as a function of the bias potential at $\hbar\omega = 2\mu$. Here we set the parameters as $\mu = 0.1$ eV, $\delta = -0.25$ eV, , $v_x = 0.1$ eV\AA, $v_y = 0.3$ eV\AA and $B=1$ eV\AA$^{2}$.}
    \label{fig:Fig4}
\end{figure*}
After performing the straightforward calculations, one can easily express the Stokes parameters in terms of the longitudinal and transverse components of the second-order conductivity and incident polarization angle as $S_0 = |\sigma^{(2)}_{\rm L}|^2+|\sigma^{(2)}_{\rm T}|^2$ and 
\begin{align}
S_1&= 
  (|\sigma^{(2)}_{\rm L}|^2-|\sigma^{(2)}_{\rm T}|^2)\cos (2\phi) 
- 2 |\sigma^{(2)}_{\rm L}||\sigma^{(2)}_{\rm T}| \cos\delta \sin (2\phi), 
\nonumber\\
S_2&= 2\sqrt{S^2_0 - S^2_1} |\sigma^{(2)}_L||\sigma^{(2)}_T|\cos\delta_0, 
\nonumber \\
S_3&= 2\sqrt{S^2_0 - S^2_1} |\sigma^{(2)}_L||\sigma^{(2)}_T|\sin\delta_0.
\end{align}
The phase difference between the longitudinal and transverse components is denoted by $\delta = \delta_L - \delta_T$. The cosine of the phase difference between  $a_x$ and $a_y$, i.e $\delta_0=\delta_y-\delta_x$, is obtained (see Appendix~\ref{sec:AppI})
\begin{align}
    \cos\delta_0 = \frac{X}{\sqrt{X^2 + Y^2}},
\end{align}
in which we define $X=-(|\sigma^{(2)}_{\rm L}|^2-|\sigma^{(2)}_{\rm T}|^2)\sin (2\phi) 
- 2 |\sigma^{(2)}_{\rm L}||\sigma^{(2)}_{\rm T}| \cos\delta \cos (2\phi)$ and $Y=2|\sigma_T^{(2)}||\sigma_L^{(2)}|\sin\delta$.
In addition to these angular quantities, the state or character of the polarization is assigned by another dimensionless parameter, known as helicity (see Appendix~\ref{sec:AppI}): 
\begin{align}
    h = \frac{|a_l|^2 - |a_r|^2}{|a_l|^2 + |a_r|^2} = \frac{2{\rm Im}[a_x a^\ast_y]}{|a_x|^2+
    |a_y|^2}=\sin(2\rchi).
\end{align}
Here, the sign of the helicity decides the left or right-handed character of the SHG signal. Specifically, if $h=+1$, the light is completely right-handed in nature while $h=-1$ refers to left-handed character of the light. Note that the helicity is proportional to the phase difference of longitudinal and transverse response, $
h\propto \sin\delta$, that can be utilized by probing the value of $\delta$. 
The presented formalism for the computation of the nonlinear polarization quantities Faraday rotation and ellipticity is general and applicable for all systems. In this work, we have discussed the case for the time-reversal symmetric and inversion broken system 1T$'$-WTe$_2$.
\section{Results and Discussion}
In this section, we present our numerical results and discuss the frequency and bias potential dependence of the nonlinear conductivity. Particularly, we report the polarization analysis by evaluating a non-vanishing Faraday rotation and helicity of SHG signal in single-layer 1T$'$-WTe$_2$. 

\subsection{Frequency dependence of the second-order susceptibility}
In Fig.~\ref{fig:Fig3}, we plot three independent tensor elements of the nonlinear susceptibility as a function of the scaled incident frequency $\hbar\omega/\mu$. The absence of inversion symmetry leads to the spin-polarized band structure, however, the second-order response function is spin degenerate due to the time-reversal invariance. Although both one and two-photon interband resonances are present, the two-photon one at $\hbar\omega/\mu=1$ is less pronounced in $\rchi^{(2)}_{yxx}$ owing to the absence of the corresponding two-photon vertex coupling $\kappa_{xy}=0$. 
The situation does not remain the same in the case of $\rchi_{xxy}^{(2)}$, and $\rchi_{yyy}^{(2)}$ where both one-photon and two-photon absorption processes contribute to the diamagnetic and paramagnetic susceptibilities. 
Further, the shape of these interband resonances can be understood through the anisotropic Fermi surface in the $k_x$-$k_y$ 2D momentum plane. The presence of the $v_y k_y$ term along the $y$-component of the Pauli basis in the modeled Hamiltonian creates the anisotropy in the Fermi surface of WTe$_2$.
On moving towards the higher momentum values, the quadratic term $Bk^2$ becomes large over the linear terms which ultimately suppress the anisotropic behavior of the Fermi surface and turns the surface into a more isotropic form at larger $k$. 
In addition to the two interband resonances, we find another peak for $\hbar\omega > 2\mu$ in Fig.~\ref{fig:Fig3}b, and Fig.~\ref{fig:Fig3}f. This arises due to the presence of the fundamental gap $2\delta$ at $\Gamma$ point ($k=0$) and for $\delta =0$, the latter peak is absent. 
In addition to these features, we check the tunability of the second-order susceptibility by altering the bias potential $U$ and we observe strong dependence on $U$ as also discussed in the next subsection.

\begin{figure*}[tp]
    \centering
    \includegraphics[width=15cm]{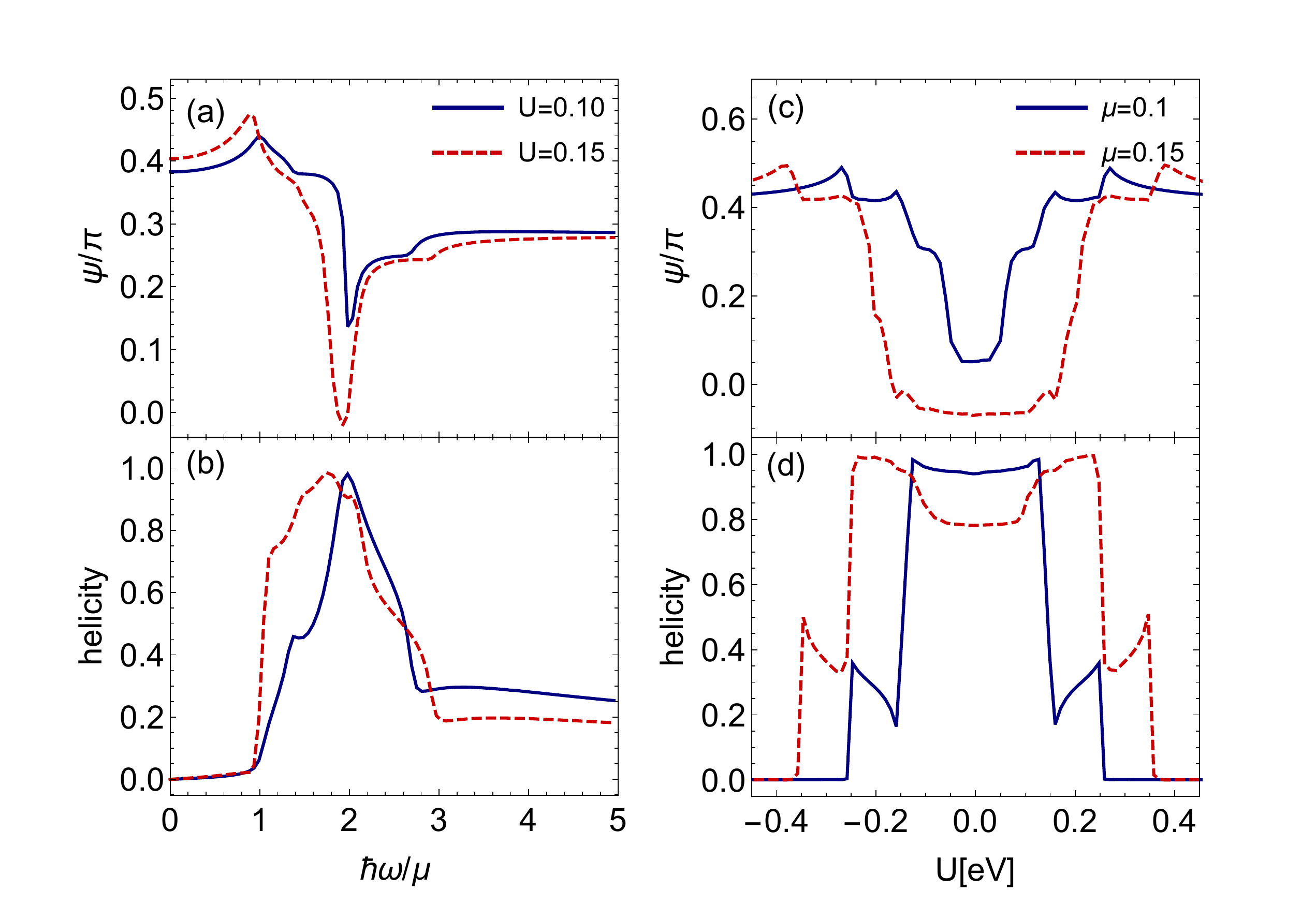}
    \caption{Faraday rotation angle and helicity (a) and (b) as a function of scaled incident energy at distinct values of $U$, but fixed $\mu=0.1$ eV and $\delta = -0.25$ eV, and Panels (c) and (d) show the variation with bias potential $U$ at the incident energy equals to the twice of the chemical potential with two different values of chemical potential, but fixed $\delta=-0.25$ eV, $B=1$ eV\AA$^{2}$, $v_x = 0.1$ eV\AA, $v_y = 0.3$ eV\AA, and the polarization angle $\phi = \pi/6$.}
    \label{fig:Fig5}
\end{figure*}

\subsection{Effect of bias potential}

To track the evolution of the nonlinear response with the out-of-plane electric field, we plot the real and imaginary parts of the second-order susceptibility versus the bias potential $U$ at the frequency $\hbar\omega = 2\mu$ where all responses contribute significantly as seen in Fig.~\ref{fig:Fig3}. We set other parameters as $\mu = 0.1$ eV, $B = 1$ eV\AA$^{2}$, $v_x = 0.1$ eV\AA, $v_y = 0.3$ eV\AA, and $\delta = -0.25$ eV for three elements $yxx$, $xxy$, and $yyy$ in Fig.~\ref{fig:Fig4}. 

In the absence of the bias potential, $U=0$, the second harmonic susceptibility vanishes owing to the preserved inversion symmetry. Thus, one requires a finite $U$ to break the inversion symmetry which gives a non-zero response. By increasing $U$, the band gap around $Q$ points widens and further enhances the susceptibility as shown in Fig.~\ref{fig:Fig4}. For small values of $U<\mu$ the response function linearly depends on the bias potential as the Fermi level lies within the conduction band. The response function changes sign when $U\to-U$ which implies that it must be an odd function of $U$. For larger $U$ but still smaller than $\hbar\omega$, we notice some resonances due to interband transitions at $\hbar\omega= 2\mu + v_x Q$. The presence of an external electric field can alter the bias potential and therefore control the value of second-order response in the distorted monolayer WTe$_2$. 

\subsection{Helicity and Faraday rotation of second harmonic signal}

Before analyzing the helicity of SHG signal in 1T$'$-WTe$_2$, it is useful to discuss it for the trigonal prismatic TMD systems such as single-layer MoS$_2$. In single-layer MoS$_2$ the helicity vanishes due to the mirror ($x\to -x$) and three-fold symmetries. Under mirror and three-fold symmetries, the only non-zero tensor elements are given by $-\sigma^{(2)}_{yyy}=\sigma^{(2)}_{xxy}=\sigma^{(2)}_{xyx}=\sigma^{(2)}_{yxx}$. Accordingly, we have $(\sigma^{(2)}_{L},\sigma^{(2)}_{T}) = -\sigma^{(2)}_{yyy} (\sin(3\phi),\cos(3\phi))$. Here, the longitudinal and transverse conductivity components are in-phase ($\delta=0$) and therefore the helicity vanishes, i.e. $h\propto\sin\delta =0$. This result implies that not every noncentrosymmetric system can generate a second harmonic signal with a finite helicity. The finite helicity is the result of chirality of the system similar to the optical activity effects \cite{barron2009molecular}.

In Fig.~\ref{fig:Fig5}a and \ref{fig:Fig5}b, we plot the Faraday rotation angle $\psi/\pi$ and helicity versus the incident energy $\hbar\omega$ at fixed $\delta = -0.25$ eV and $\mu=0.1$ eV. In general, the Faraday rotation and helicity depend on the incident polarization angle $\phi$, see for instance  Eq.~\eqref{eqn:FR}. Here, we consider the polarization angle $\phi = \pi/6$, thus the value of $|\psi |$ for $\hbar\omega < \mu$ approaches to an asymptotic finite value due to the presence of a nonlinear transverse current component. At low excitation energy $\hbar\omega \le \mu$, the real part of the nonlinear response function goes to zero which implies a vanishing helicity since $a_{x,y}$ are purely real-valued. 

At $\hbar\omega>\mu$, both the real and imaginary parts of the second order response are finite, as shown Fig.~\ref{fig:Fig3}, because the interband transitions lead to a non-vanishing helicity. This is a signature of the conversion of the linearly polarized light into the right-handed polarization with double frequency. As the energy approaches twice the chemical potential, the system shows maximum positive helicity. At extreme higher energies, the second-order conductivity changes monotonically by $\hbar\omega$ which ultimately yields a saturation behavior in the Faraday angle and the helicity in the high energy regime. In addition to the energy variation of the quantities, we also observe that the increase in the bias potential displaces the overall behavior of the polarization quantities towards the higher energy values due to the significant variations in the band structure of 1T$'$-WTe$_2$ as discussed earlier in the case of second-order response.

To elaborate the bias potential effect, we show the results for the Faraday angle and helicity of 1T$'$-WTe$_2$ at one-photon resonance frequency $\hbar\omega = 2\mu$ in Fig.~\ref{fig:Fig5}c and Fig.~\ref{fig:Fig5}d that corresponds to the maximum light conversion to the right-handed state of the polarization. 
For a given chemical potential and one-photon resonance condition for the frequency, we can notify a sudden jump from vanishing helicity to a finite value at $\Delta_{-}= U-v_x Q=2\mu$ for $U>0$. Further decrease of the bias potential by the vertical electric field leads to another jump to a larger value of the helicity at $\Delta_{-}= U-v_x Q=\mu$. For a wide range around of small $U$ the helicity is almost full $h\sim 1$. The corresponding kinks are also visible in the Faraday rotation angle plot versus the bias potential. By moving the Fermi level within the conduction band to higher energy, the number of available states for conduction increases. This ultimately shifts the helicity jump locations to higher values $\mu+ v_xQ$ and $2\mu+ v_xQ$. Since the interband transitions are not allowed at $ |U| > 2\mu$ for $\hbar\omega=2\mu$, the helicity vanishes for large bias potential as shown in Fig. \ref{fig:Fig5}d. Furthermore, the behavior of the Faraday rotation angle and helicity remain symmetrical on interchanging the sign of the bias potential.

Finally, we refer to the experimental setup \cite{xu_NP2018} for the dual gated device of the encapsulated single-layer WTe$_2$ between thin layers of hexagonal boron nitride (hBN) that demonstrate the tunable potential by an external vertical electric field. The corresponding electrical gating results in the net displacement electrical field (or bias potential) between the top and bottom atomic layers which breaks the inversion symmetry. The bias potential is determined by the relation $U \approx e \epsilon E_z d$, where $\epsilon$ is the dielectric constant of hBN, $E_z$ is the external vertical electric field, and $d$ is the separation between the top and bottom atomic layers of 1T$'$-WTe$_2$. By setting $E_z \approx 0.1$V/nm, $\epsilon \approx 3$, and $d \approx 0.31$ nm \cite{akash_NPJ2018, shah_PRB2019}, the above formula implies $U\sim 0.1$ eV where we have theoretically observed the maximum susceptibility and the highest second harmonic helicity.
\section{Summary}
To summarize, we studied the Faraday rotation angle and helicity for the second harmonic radiation in an inversion symmetry broken, distorted and gated 2D single-layer 1T$'$-WTe$_2$. To compute the second harmonic helicity, we first calculate second-order conductivity within a diagrammatic framework. We found that the linearly polarized light shows maximum conversion into a second harmonic signal with right or left-handed helicity on propagating through the system when the incident energy is in resonance with the interband transition edges. We identified that the linear-to-circular conversion is pronounced only within the window of two-photon transition-edge due to the encapsulated features by the band structure of the system. 
Our results provide a pave to the future helicity spectroscopy experiments which may help to probe hidden topology in the nonlinear spectroscopy. 

\section*{Acknowledgments} 
This work is supported by Nordita and the Swedish Research Council (VR 2018-04252). We thank G. Soavi for carefully reading the manuscript and his very useful comments. 

\bibliography{refs}
 
\onecolumngrid
\appendix
\section{Derivation of Faraday rotation angle and helicity}
\label{sec:AppI}
Consider a linearly polarized incident beam of light striking at the 2D material interface $z=0$ in the $xy$-plane which is defined as
\begin{align}
\bm{E} = \hat{\bm \epsilon}({\phi}) E_0 e^{i({\bf q}\cdot {\bf r} - \omega t)} +c.c,
\end{align}
where the polarization unit vector is $\hat{\bm \epsilon}({\phi}) = \hat {\bf x} \cos\phi + \hat{\bf y} \sin\phi $ and the photon wave vector ${\bf q}$, having $\phi$ the polarization angle with $\hat{x}$-direction, $\omega$ the frequency of the incoming beam, $E_0$ the amplitude of the beam, $t$ refers to time. By definition, the second harmonic current reads 
\begin{align}
J^{(2)}_a(2\omega) = \sigma^{(2)}_{abc}(2\omega;\omega,\omega) E_b(\omega) E_c(\omega).
\end{align}
In terms of the longitudinal and transverse components, the second-order current can be decomposed as follows 
\begin{align}
{\bf J}^{(2)} =  J^{(2)}_{\rm L} \hat{\bm \epsilon}(\phi) + J^{(2)}_{\rm T} \hat {\bf t}(\phi),
\end{align}
Note the transverse unit vector $\hat {\bf t}(\phi) = \hat {\bf z}\times \hat{\bm \epsilon}(\phi)= \hat{\bf y}\cos\phi -\hat{\bf x}\sin\phi$. We define the nonlinear conductivity associated with the longitudinal and transverse directions: 
\begin{align} \label{eqn:Axy}
\sigma^{(2)}_{\rm L} = {J^{(2)}_{\rm L}}/{{\cal E}^2_0} = a_x \cos\phi + a_y \sin\phi, \quad 
\sigma^{(2)}_{\rm T}= {J^{(2)}_{\rm T}}/{{\cal E}^2_0} = -a_x\sin\phi + a_y\cos\phi.
\end{align}
Here, the coefficients $a_x$, and $a_y$ can be calculated by writing $a_x = |a_x| e^{i\delta_x}$, and $a_y = |a_y| e^{i\delta_y}$. From Eq.~\eqref{eqn:Axy}, we have
\begin{align}
a_x = \sigma_{\rm L}^{(2)} \cos\phi - \sigma_{\rm T}^{(2)} \sin\phi = |\sigma^{(2)}_{\rm L}|e^{i\delta_{\rm L}} \cos\phi - |\sigma^{(2)}_{\rm T}| e^{i\delta_{\rm T}} \sin\phi\\
a_y = \sigma_{\rm L}^{(2)} \sin\phi + \sigma_{\rm T}^{(2)} \cos\phi = |\sigma^{(2)}_{\rm L}|e^{i\delta_{\rm L}} \sin\phi + |\sigma^{(2)}_{\rm T}| e^{i\delta_{\rm T}} \cos\phi.
\end{align}
Straightforward algebraic calculation yields the norm of $a_x$ and $a_y$ like
\begin{align}
&|a_x| = \sqrt{ |\sigma^{(2)}_{\rm L}|^2 \cos^2\phi + |\sigma^{(2)}_{\rm T}|^2 \sin^2\phi - 
|\sigma^{(2)}_{\rm L}| |\sigma^{(2)}_{\rm T}| \cos(\delta) \sin(2\phi)}
\\
&|a_y| = \sqrt{ |\sigma^{(2)}_{\rm L}|^2 \sin^2\phi + |\sigma^{(2)}_{\rm T}|^2 \cos^2\phi +
|\sigma^{(2)}_{\rm L}| |\sigma^{(2)}_{\rm T}| \cos(\delta)  \sin(2\phi)}.
\end{align}
and similarly the phase difference is given by 
 
\begin{align} \label{eqn:tand0}
&\tan(\delta_0) =   \frac{2|\sigma^{(2)}_{\rm T}||\sigma^{(2)}_{\rm L}|\sin(\delta)}{2|\sigma^{(2)}_{\rm T}||\sigma^{(2)}_{\rm L}|\cos(\delta) \cos2\phi+
(|\sigma^{(2)}_{\rm T}|^2 - |\sigma^{(2)}_{\rm L}|^2)\sin2\phi}. 
\end{align}
where $\delta_0 = \delta_y - \delta_x$ and $\delta = \delta_{\rm L} - \delta_{\rm T}$. 
%
Stokes parameters for the monochromatic light consist of four scalar values  related to the polarization angles \cite{mcmaster_AJP1954, jackson_book, artal_book},
\begin{align}
& S_0 = |a_x|^2 + |a_y|^2\\ \nonumber 
& S_1 = |a_x|^2 - |a_y|^2 = S_0 \cos(2\rchi) \cos(2\psi) \\ \nonumber
& S_2 = 2|a_x| |a_y| \cos\delta_0 = S_0 \cos(2\rchi) \sin(2\psi) \\ \nonumber
& S_3 = 2|a_x| |a_y| \sin\delta_0 = S_0 \sin(2\rchi).
\end{align}
Here $S_0$ represents the irridiance of the light beam, $S_1$ denotes the dominant character of the horizontal and vertical components based on the sign, $S_2$ refers to the orientation of the ellipse, $S_3$ corresponds to the handedness of the polarization state, $\psi$ is the angle of rotation made by polarized light with respect to the $x$-axis or the original semi-major axis, and $\rchi$ is the ellpticity that define the angular amount of ellipse from the circular shape. Similarly, in the circular basis $(\hat{l},\hat{r})$ we can define these parameters below
\begin{align}
& S_0 = \vert a_l\vert^2 + \vert a_r\vert^2\\ \nonumber 
& S_1 = 2\text{Re} (a_l^* a_r) \\ \nonumber
& S_2 = -2\text{Im} (a_l^* a_r)\\ \nonumber
& S_3 = \vert a_l\vert^2 - \vert a_r\vert^2.
\end{align}
Now, on dividing $S_2$ by $S_1$, the Faraday rotation angle can be obtained as
\begin{align}
\tan (2\psi) = \frac{S_2}{S_1} = \frac{ 2|a_x| |a_y| \cos\delta_0}{|a_x|^2 - |a_y|^2}.
\end{align}
 Similarly, the ellipticity can be calculated using the relation
\begin{align}\label{eq:ellipticity}
\tan (2\rchi) = \frac{S_3}{\sqrt{S_1^2 + S_2^2}} = \tan\delta_0 \sin (2\psi).
\end{align}
The helicity which is the ratio of the handedness of the polarization state and the  irridance of the light beam is defined like
\begin{align}
    h  = \frac{|a_l|^2 - |a_r|^2}{|a_l|^2 + |a_r|^2} = \frac{S_3}{S_0}.
\end{align}
Using the relation for the ellipticity, the helicity becomes

\begin{align}
    h = \sin (2\rchi).
\end{align}
\end{document}